\documentclass[notitlepage,12pt]{article}

\usepackage{amssymb}
\begin{document}
\makeatletter \@addtoreset{equation}{section}
\pagenumbering{arabic}
%\pagestyle{myheadings}

%\begin{titlepage}

\title{Ground state and functional integral representations  of the CCR algebra \\ with free
evolution}
\author{J. L\"{o}ffelholz,
%G. Morchio, F. Strocchi}
\\  Berufsakademie, Leipzig  \and
G. Morchio,
\\ Dipartimento di Fisica, Universita' di Pisa and INFN, Pisa \and
F. Strocchi
\\  Scuola Normale Superiore and INFN, Pisa}

\date{}
\maketitle

\begin{abstract}

The problem of existence of ground state representations of the
CCR algebra with free evolution are discussed and all the
solutions are classified in terms of non regular or indefinite
invariant functionals. In both cases one meets unusual
mathematical structures which appear as prototypes of phenomena
typical of gauge quantum field theory, in particular of the
temporal gauge. The functional integral representation in the
positive non regular case is discussed in terms of a generalized
stochastic process satisfying the Markov property. In the
indefinite case the unique time translation and scale invariant
Gaussian state is suprisingly faithful and its GNS representation
is characterized in terms of a Tomita-Takesaki operator. In the
corresponding Euclidean formulation, one has a generalization of
the Osterwalder-Schrader reconstruction and the indefinite Nelson
space, defined by the Schwinger functions, has a unique Krein
structure, allowing for the construction of Nelson projections,
which satisfy the Markov property. Even if Nelson positivity is
lost, a functional integral representation of the Schwinger
functions exists in terms of a Wiener random variable  and a
Gaussian complex variable.
\end{abstract}

%\end{titlepage}

%\begin{document}
%\textwidth=6in \textheight=8.5in \pagestyle{myheadings}
\makeatletter \@addtoreset{equation}{section}
\pagestyle{myheadings}

\newtheorem{Theorem}{Theorem}[section]
\newtheorem{Definition}{Definition}[section]
\newtheorem{Proposition}{Proposition}[section]
%\numberwithin{equation}{section}

%\include{defu}
\def \be {\begin{equation}}
\def \ee {\end{equation}}
\def \ume {{\scriptstyle{\frac{1}{2}}}}
\def \ra {\rightarrow}
\def \Ra {\Rightarrow}
\def \eqq {\equiv}

\def \a {{\alpha}}
\def \b {{\beta}}
\def \g {{\gamma}}
\def \d {{\delta}}
\def \eps {{\varepsilon}}
\def \th {{\theta}}
\def \l {{\lambda}}
\def \La {{\Lambda}}
\def \s {{\sigma}}
\def \Si {{\Sigma}}
\def \t {{\tau}}
\def \ph {{\varphi}}
\def \phb {{\overline{\varphi}}}
\def \o {{\omega}}
\def \Om {\mbox{${\Omega}$}}
\def \Ga {{\Gamma}}

\def \A {{\cal A}}
\def \B {{\cal B}}
\def \C {{\cal C}}
\def \D {{\cal D}}
\def \F {{\cal F}}
\def \G {{\cal G}}
\def \H {\mbox{${\cal H}$}}
\def \J {{\cal J}}
\def \K {{\cal K}}
\def \L {{\cal L}}
\def \N {{\cal N}}
\def \O {{\cal O}}
\def \P {{\cal P}}
\def \S {{\cal S}}
\def \U {{\cal U}}
\def \V {{\cal V}}
\def \W {{\cal W}}
\def \Z {{\cal Z}}

\def \Id {\mathop{\bf Id}}
\def \id {{\bf 1 }}
\def \eijk {{\varepsilon_{ijk}}}
\def \eklm {{\varepsilon_{klm}}}
\def \dij {{\delta_{ij}}}
\def \dik {{\delta_{ik}}}
\def \djk {{\delta_{jk}}}
\def \dkl {{\delta_{kl}}}
\def \Psio {{\Psi_0}}

\def \di {{\partial_i}}
\def \dj {{\partial_j}}
\def \dk {{\partial_k}}
\def \dl {{\partial_l}}
\def \do {{\partial_0}}
\def \dz {{\partial_z}}

\def \dmu {{\partial_\mu}}
\def \dnu {{\partial_\nu}}
\def \dla {{\partial_\lambda}}
\def \dr {{\partial_\rho}}
\def \ds {{\partial_\sigma}}
\def \dt {{\partial_t}}
\def \do {{\partial_0}}
\def \dum {{\partial^\mu}}
\def \dun {{\partial^\nu}}
\def \Amu {{A_\mu}}
\def \Anu {{A_\nu}}
\def \Fmn {{F_{\mu\,\nu}}}
\def \jm  {j_\mu}

\def \abf {{\bf a}}
\def \bbf {{\bf b}}
\def \cbf {{\bf c}}
\def \hbf {{\bf h}}
\def \k {{\bf k}}
\def \kbf {{\bf k}}
\def \jbf {{\bf j}}
\def \j   {{\bf j}}
\def \nbf {{\bf n}}
\def \q {{\bf q}}
\def \qbf {{\bf q}}
\def \p {{\bf p}}
\def \pbf {{\bf p}}
\def \sbf {{\bf s}}
\def \rbf {{\bf r}}
\def \ubf {{\bf u}}
\def \vbf {{\bf v}}
\def \xbf {{\bf x}}
\def \x {{\bf x}}
\def \y {{\bf y}}
\def \ybf {{\bf y}}
\def \v {{\bf v}}
\def \z {{\bf z}}
\def \zbf {{\bf z}}
\def \Rbf {{\bf R}}
\def \Cbf {{\bf C}}
\def \Nbf {{\bf N}}
\def \Zbf {{\bf Z}}
\def \Abf {{\bf A}}
\def \Jbf {{\bf J}}

\newcommand{\mbf}[1] {\mbox{\boldmath{$#1$}}}

\def \AO {{\cal A}({\cal O})}
\def \AO' {{\cal A}({\cal O}')}
\def \Aob {\A_{obs}}
\def \dxy {\delta(x-y)}
\def \at {{\alpha_t}}
\def \ax {{\alpha_{\x}}}
\def \atv {{\alpha_t^V}}

\def \fR {{f_R}}
\def \hf {\tilde{f}}
\def \tilf {\tilde{f}}
\def \tilg {\tilde{g}}
\def \tilh {\tilde{h}}
\def \tilF {\tilde{F}}
\def \tilJ {\tilde{J}}

\def \cc  {\subseteq}
\def \] {\supseteq}

\def \pio {{\pi_\o}}
\def \pom {{\pi_{\Omega}}}
\def \Hom { {\H_{\Omega}} }
\def \Psiom  { \Psi_{\Omega} }

\def \Pf {{\bf Proof.\,\,}}
\def \limx {{\lim_{|\x| \ra \infty}}}
\def \frx {f_R(x)}
\def \limR {\lim_{R \ra \infty}}
\def \limV {\lim_{V \ra \infty}}
\def \limko {\lim_{k \ra 0}}

\def \Roo {R \ra \infty}
\def \ko {k \ra 0}
\def \jo {j_0}
\def \su {{ \left(
\begin{array}{clcr} 0 & 1 \\1 & 0 \end{array} \right)}}
\def \sd {{ \left(
\begin{array}{clcr} 0 & -i \\i & 0 \end{array} \right)}}
\def \st {{ \left(
\begin{array}{clcr} 1 & 0 \\0 & -1 \end{array} \right)}}

\def \eqq {\equiv}
\def \Pf {{\bf Proof}\,\,\,\,}
\def \Rbf {{\bf R}}
\def\Ra{\Rightarrow}
\def\ra{\rightarrow}
\def\Id{\mathop{\bf Id}}
\def\dalamb{{\sqro86}{\hskip1.1pt}}

\@addtoreset{equation}{section}
\def\theequation{\thesection.\arabic{equation}}

\def\AO'{\mbox{${\cal A}({\cal O}'$}}
\def\O{\mbox{${\cal O}$}}
\def \cal {\mathcal}
\def \o {{\omega}}
\def \O {{\mathcal O}}
\def \OM {{\Omega}}
\def \A {{\mathcal A}}
\def \AO {\A(\O)}
\def \B {{\cal B}}
\def \F {{\cal F}}
\def \D {{\cal D}}
\def \H {{\cal H}}
\def \K {{\cal K}}
\def \P {{\cal P}}
\def \S {{\cal S}}
\def \Sp {{{\cal S}^\prime}}
\def \Z  {{\cal Z}}
\def \W  {{\cal W}}

\def \psz {{\Psi_0}}
\def \psb {{\bar\psi}}
\def \a {\alpha}
\def \b {\beta}
\def \d {\delta}
\def \l {\lambda}
\def \t {\tau}
\def \g {{\gamma}}
\def \Cf {{C^\infty}}

\def\naturali{{\bf N}}
\def \rep {representation }
\def \reps {representations }
\def \ip {inner product }
\def \nd {non--degenerate }
\def \qft {quantum field theory }
\def \ex {extension }
\def \exs {extensions }
\def \wcs {weakly compatible sequences }
\def \ucs {u-weakly compatible sequences }

\def \eps {{\varepsilon}}
\def \Dz {{{\cal D}_0}}
\def \Da {{{\cal D}_\alpha}}
\def \Iz {{{\cal I}_0}}
\def \Iza {{{\cal I}_0^\alpha}}
\def \deg {<x,y> \ = 0, \ \ \forall y \in \Dz}
\def \fayz {\forall y \in \Dz}
\def \tw {{\tau_w}}
\def \pyx {{{p_y (x)}}}
\def \ps {{ < {\cdot} , {\cdot} > }}
\def \psnm {{ < x_n , y_m > }}
\def \xn { \{ x_n \} }
\def \yn { \{ y_n \} }
\def \xnx {{x_n \to x}}
\def \yny {{y_n \to y}}
\def \lnm {{\lim_n \lim_m}}
\def \lmn {{\lim_m \lim_n}}
\def \LA  {\Lambda}
\def \Ta {{T_\alpha }}
\def \Sa {{S_\alpha }}
\def \ux {{\underline x}}
\def \uy {{\underline y}}
\def \uz {{\underline z}}
\def \uf {{\underline f}}
\def \xnDz {{x_n \in \Dz}}
\def \dm {{\partial_\mu}}
\def \dn {{\partial_\nu}}

\def \di {{\partial_i}}
\def \dj {{\partial_j}}
\def \dk {{\partial_k}}
\def \dt {{\partial_t}}
\def \de {{\partial}}

\def \be {\begin{equation}}
\def \ee {\end{equation}}
\def \e {\end}
\def \yyyyy {\end{equation}}
\def \x {{\bf x}}
\def \y {{\bf y}}
\def \z {{\bf z}}
\def \k {{\bf k}}
\def \SR {\S(\Rbf^3)}
\def \DR {\D(\Rbf^3)}
\def \psio {\Psi_0}
\def \reali {\Rbf}
\def \complessi {{\bf C}}

\def \AV {{{\cal A}_V}}
\def \Ut {{{\cal U}(t)}}
\def \Ha {{{\cal A}_H}}
\def \Wa {{{\cal A}_W}}
\def \Hap {{{\cal A}_H^\prime}}
\def \o {{\omega}}
\def \O {{\Omega}}
\def \ni {\noindent}
\def \psV {{\Psi_V}}
\def \pioH {{\pi_\o (\Ha)}}
\def \pioHp {{\pi_\o (\Ha)^\prime}}

\section{Introduction}
Most of the wisdom on quantum field theory and more generally on
quantum systems with infinite degrees of freedom relies on the
existence of a vacuum or ground state, so that their formulation
and control is obtained in terms of the ground state correlation
functions. In particular the analytic continuation to imaginary
time and the functional integral approach exploits this basic
property ~\cite{STW, GJ}. For a large class of models involving
infrared singular canonical variables or fields the existence of a
ground state is linked to non regular representations of such
variables ~\cite{AMS1,AMS2,LMS1,LMS2} if positivity is required,
whereas regular, i.e. weakly continuous, representations are
available if one allows indefinite metric ~\cite{MPS,MPS1, S}.

The mathematical problems emerging in these cases, including the
euclidean functional integral formulation, already show up if one
ask for the simple case of the ground state representations of the
Heisenberg algebra with free time evolution. In particular the
model reproduces the basic properties of the temporal gauge in
quantum electrodynamics ~\cite{CT}, where the longitudinal
variables $div {\bf A}, \,\,div {\bf E}$ correspond to a
collection of free quantum mechanical canonical variables
~\cite{LMS3}.

Thus the analysis of the ground state representations of the model
is an essential step for the solution of mathematical problems and
for the construction of functional integral representations of the
temporal gauge; in particular the model also displays the
occurrence of unusual features in the temporal gauge, including
the conflict between time translation invariance, energy spectral
condition, positivity and the regularity of the vacuum state
~\cite{LMS3}. One also obtains non trivial information on the
functional integral formulation  of models with infrared singular
variables, as in the temporal gauge. For these reasons, the
analysis of the model has  general implications on realistic
quantum field theories and should not be dismissed on the basis of
its trivial dynamics.

In Sect.2 we discuss the problem of  ground state representations
of the Heisenberg algebra, which correspond to the standard field
quantization of the temporal gauge in terms of vacuum expectation
of the canonical fields. Such representations are shown to violate
positivity  and share the basic features of the Tomita-Takesaki
theory  ~\cite{BR, H}. The unique time translation and scale
invariant state is in fact faithful on the Heisenberg algebra and
a Tomita-Takesaki operator $S$ is defined by $ S A \Psio = A^*
\Psio$. The corresponding GNS representation, with cyclic ground
state vector $\Psio$, is given by the tensor product of a Fock and
anti-Fock representations ~\cite{MSV, MS} of the canonical
variables $Q_\pm \eqq (q \pm p')/\sqrt{2},\,\,\,P_\pm \eqq (\pm p
+ q')/\sqrt{2}$, with $q'\eqq J q J^*,\,\,\,  p'\eqq J p J^* $,
where $J$ is an antiunitary operator obtained from $S$ as in the
Tomita-Takesaki theory. This implies  that the non positive vacuum
of the (standard) temporal gauge has  properties similar to those
 of a KMS state.

The ground state positive representations of the Weyl algebra are
non regular and satisfy the energy spectral condition. The
euclidean version satisfies Osterwalder-Schrader positivity and
the corresponding functional integral representation satisfies
Markov property  ~\cite{GJ,N}. Actually, the euclidean correlation
functions can be expressed as expectations of the euclidean fields
$U(\a, \t) = e^{i\a x(\t)}$ with a functional measure which is the
product of the conditional Wiener measure $d W_{0, x}(y(\t))$ on
trajectories $y(\t)$ starting at $x$ at time $\t=0$, times the
ergodic mean $d \nu(x)$, which defines a measure on the Gelfand
spectrum of the Bohr algebra ~\cite{LMS1}  generated by the $U(\a,
\t)$. The Osterwalder-Schrader reconstruction theorem provides the
(non regular) quantum mechanical representation, with a ground
state, which is cyclic with respect to the euclidean algebra at
time zero.

In Sect. 3 we analyze  the euclidean correlation functions of the
indefinite representation of the Heisenberg algebra. We show that
the failure of the Ostervalder-Schrader positivity does not
prevent an Osterwalder-Schrader reconstruction, yielding the
(indefinite) quantum mechanical representation. Nelson positivity
fails and we have an indefinite inner product space with a Krein
structure very similar to that of the massless scalar field in two
space time dimensions ~\cite{MPS}. A generalization of Nelson
strategy ~\cite{N} can be performed and projection operators
$E_{\pm}, E_0$ can be defined, yielding the same operator
formulation of the Markov property as in the positive case.

A stochastic interpretation of the euclidean variables meets the
problem of absence of Nelson positivity and of the non existence
of invariant measures for the brownian motion. It is shown that a
functional integral representation of the euclidean (indefinite)
correlation functions $< x(\t_1) ...x(\t_n)>$ exists in terms of a
Brownian variable $\xi(\t)$ and of complex variables $z,\, \bar{z}
$ $$< x(\t_1) ... x(\t_n)>\, =$$ $$ \int  d W_{0,0}(\xi(\t))\, d z
\,d \bar{z}\,e^{-2 |z|^2}\, (\xi(\t_1) + z - |\t_1|
\bar{z})...(\xi(\t_n) + z - |\t_n| \bar{z}).$$

The above analysis provides a framework which can be applied e.g.
to the massless scalar model in two space time dimensions to
obtain a functional integral representation of its euclidean
correlation functions.

%%%%%%%%%%%%%%%%%2222222222222222222222222222222222222%%%%%%%%%%%
\section{Ground state representations of the Heisenberg algebra
with free evolution}

Our model  is defined by the Heisenberg *-algebra $\Ha$ generated
by $q, p$ (for simplicity we consider the one dimensional case),
invariant under the *-operation (hermiticity) and satisfying the
canonical commutation relations (CCR) \be{[\, q ,\, p \,] = i.}
\ee invariant under the *-operation $q = q^*$, $ p = p^*$ The time
evolution is defined by the following one parameter group of
*-automorphisms $\alpha_t$, $t \in \reali$ \be \alpha_t (q) = q +
p \,t \equiv q(t) \ , \ \ \ \ \ \ \alpha_t (p) = p \ . \ee
Motivated by Wightman formulation of quantum field theory, it is
natural to ask whether there are time translationally invariant
hermitean linear functionals $\o$ on $\Ha$, $ \o (\alpha_t (A)) =
\o (A)$, and to investigate the generalized GNS representations
defined by them.

By eq.\,(2.2) the correlation functions
$\omega(A\,\alpha_t(B)\,C)$, $A, B, C, \in \A_H$ are polynomials
in $t$ and therefore time translation invariance of $\omega$
implies that their Fourier transform has support at the origin and
therefore satisfy the energy spectral condition, which
characterizes ground states.
\begin{Proposition} A hermitean linear functional $\o$  on the Heisenberg algebra
invariant under time translations, i.e. such that $\forall A \in
\Ha$, $ \o (\alpha_t (A)) = \o (A)$,    is not positive, defines a
GNS representation in an indefinite inner product space $D =
\pi_\omega (\Ha) \Psi_{\o}$ and the two point functions have the
following form
\be
\o(q^2) = c, \, \ \ \ \ \o(p^2) = 0, \ , \ \ \ \ \o (q p) = - \o(p
q) = i/2 \ , \ee where $c$ is a constant.

If $\o$ is Gaussian, i.e. its truncated correlation functions
vanish, the GNS representation $(D,\, \pi_\o)$ defined by it has a
non trivial commutant $\pi_\o(\A_H)^\prime$.

Moreover, if $\o$ is Gaussian and $c = 0$, then $\o$ is invariant
under scale transformations $$ q \ra \l\,q, \,\,\,\,\,p \ra
\l^{-1}\,p, \,\,\,\,\l \in \Rbf,$$ and such transformations are
implemented by isometric operators in $D$.

\end{Proposition}
\Pf \,\,\, Since $$ \o(p^2) = \o (d \alpha_t(q) / d t \;\, p) =
d/dt \; \o (\alpha_t (q p)) = 0, $$  if positivity holds, $
|\omega([ q , \, p ])| \leq \omega(q^2)^{1/2} \omega(p^2)^{1/2} =
0 $, in contrast with the CCR.

Equations (2.3) follow from the condition of time translation
invariance of the two point function $\o (q(t) \,q(s))$ and the
CCR.

By a standard GNS construction $\o$ provides a representation
$\pi_\o$ of the Heisenberg algebra on a vector space $D =
\pi_\omega (\Ha) \Psi_\o$, with an indefinite non-degenerate inner
product given by $$< A \Psi_\o, \, B \Psi_\o > = \o( A^* \,B ).$$
If $\o$ is Gaussian, the matrix elements of the operator $ U(t) $
which implements the time translations on $ D $ can be explicitly
computed and the (weak) derivative $ d U(t)/ d t $ exists and
defines a hermitean operator $ H $ on $ D $. Without loss of
generality, we can redefine $ H $ so that $ H \Psi_\o = 0$;
henceforth $\Psi_\o$ will be denoted by $\psz$ . Now, the CCR
imply
\be
H = p^2 /2 + h \ , \ \ \ \ h \in \pioHp \ee and, if $\pioHp = \{
\lambda {\bf 1} \} \ , \lambda \in \complessi$, the equations
$(\psz , H \psz) = 0$, $\o(p^2)=0 $ imply $h = 0$. This leads to a
contradiction $$ 0 = \o (q^2 H) = \o (q p)^2 = -1/4.$$ The last
statement follows from scale invariance of the correlation
functions. $\blacksquare$

 \vspace{2mm} The existence of a non-trivial commutant
is reminiscent of the structure which characterizes the theory of
KMS (non zero temperature) states and more generally the
Tomita-Takesaki theory. The occurrence of such a structure in a
ground state representation is a consequence of the fact that the
functionals of Proposition 2.1 are faithful.

\begin{Proposition}  The Gaussian functionals $\o$ of Proposition 2.1 are
faithful, i.e.,
\be
\o (B A) = 0, \ \ \  \forall B \in \Ha  \ee implies $A = 0$.

Furthermore the commutant $\pi_\o(\A_H)^\prime$ is given by \be{
\pi_\o(\A_H)^\prime = S\,\pi_\o(\A_H)\,S,}\ee where $S$ is a
Tomita-Takesaki antilinear operator defined by \be{ S\, A\, \psz
\equiv A^*\, \psz, \,\,\,\,\,\forall A \in \pioH, }\ee with $\psz$
the vector which represents the state $\o$.

\end{Proposition}
\Pf \,\,Any $A \in \Ha$ can be written in an unique way as an
ordered polynomial $A = \sum_{k = 0}^{N}\, q^k \,P_k(p)$, with
$P_k$ a polynomial. Then, by time translation invariance of $\o$,
eq.(2.5) implies $\o(B \,\alpha_t(A)) = 0, \,\,\, \forall t$. Thus
the term of highest power of $t$, $\o(B\, p^N\,P_N(p))$ vanishes
$\forall B \in \Ha$; in particular $\o(q^k\,p^N\,P_N(p)) = 0, \,
\forall k$ which implies $P_N(p) = 0$, by the gaussian property of
$\o$ and eqs.(2.3).

Since $\o$ is faithful,  eq.(2.7) defines  the analogue of the
Tomita-Takesaki antilinear operator $S$ on $D$. Clearly, $ S \psz
= \psz$ and $ S^2 = 1$. As in the standard Tomita-Takesaki theory
one has $$ S A S \in \pioHp \ , \ \ \ \forall A\in \pioH. $$
Furthermore, since $\psz$ is cyclic for $\pioH$ and therefore
separating for $\pioHp$, $\forall B \in \pioHp$, $\exists A \in
\pioH$ such that $B \,\psz = A^*\,\psz$ and therefore $ B = S A
S$; in conclusion \be{ S \pioH S = \pioHp.}\ee $\blacksquare$

For $c =0$ the Tomita-takesaki structure is particularly simple
and close to the positive case; the case $c \neq 0$ requires a
more general analysis of Gaussian indefinite functionals, which
will be done elsewhere.
\begin{Proposition}  If $\o$ is the Gaussian  hermitean linear functional
on the Heisenberg algebra $\Ha$, defined by eqs.(2.3) with $c =
0$, then $ \pioHp $ is the (concrete) algebra $\Hap$ generated by
pseudo-canonical variables $q', p'$  \be{ q' \eqq  \,J\,q\,J^*,
\,\,\,\,p' \eqq \,J\,p\,J^*, \,\,\,\,\,\, [\, q' ,\, p' \,] = -i,}
\ee where $J$ is an anti-unitary operator with respect to the
indefinite inner product of $D$: \be{J\,J^* = J^* \,J = 1.}\ee J
is  related to $S$ and to the (non positive) modular operator
$\triangle \eqq S^* \,S$ by \be{S = J\,\triangle^{1/2}.}\ee

$\pi_\o(\Ha)$ is the tensor product of a Fock and an anti-Fock
representation, respectively, for the pairs of canonical variables
\be
Q_\pm  \equiv (q \pm p')/ \sqrt 2 \ , \ \ \ P_\pm  \equiv (\pm p +
q')/ \sqrt 2 \ee which satisfy \be [Q_\pm , P_\pm] = \pm i \ , \ \
\ \ [Q_\mp , P_\pm] = 0. \ee In fact,   by introducing the
 operators
\be
a \equiv (Q_{+} + i P_{+}) / \sqrt 2 \ , \ \ \ \ b^* \equiv (Q_{-}
+ i P_{-}) / \sqrt 2 \ , \ee which satisfy
\be
[\,a,\, a^*\,] = 1 =  [\,b, b^*\,], \ , \ \ \ [\,a,\,b]= [\,a,\,
b^*] = 0 \ , \ee one has that $\psz$ satisfies the Fock and
anti-Fock conditions
\be
a\, \psz = 0,  \ \ \ \ b^*\, \psz = 0. \ee

 \end{Proposition}
\Pf\,\,\,\, The adjoint of  $S$ satisfies \be{
 S^* \psz = \psz\ , \ \ \ S^{* \, 2} = 1 \ ,
\ \ \ S^* \pioHp S^* = \pioH.} \ee In fact,  one has $$ < A \psz,
S^* \psz > = \overline{ < S A \psz , \psz >} = < A \psz, \psz >;
$$ the last eq.(2.17) follows from eq.(2.8).

Because of the indefinite inner product, the analog of the modular
operator   $ \triangle \equiv S^* S $ is not positive and in fact
it is  given by $$ S^* S p^k q^j \psz = S^* q^j p^k S^* \psz =
(-1)^{j+k} p^k q^j \psz. $$ Even if $\triangle $ is not positive
one may introduce a hermitean  square root of $\triangle $ \be{
 \triangle^{1/2} p^k q^j \psz  \equiv  i^k (-i)^j  p^k q^j \psz.
}\ee In fact  $$
 < \triangle^{1/2} p^l q^m \psz ,  \,p^k q^j \psz > =
 (-i)^l i^m \, \o ( q^m p^l p^k q^j ) =$$
 $$i^k (-i)^j \, \o ( q^m p^l p^k q^j ) =
 <\, p^l q^m \psz , \, \triangle^{1/2} p^k q^j \psz > \ ,
$$ since the above correlation functions vanish unless $ k + l = m
+ j$. A hermitean $\triangle^{-1/2}$  is defined by changing $i$
into $ -i$ in the definition of $ \triangle^{1/2} $.

In analogy with the non zero temperature  case, one may then
introduce the analog of the modular conjugation  $ J $ defined by
$J = S\, \triangle^{-1/2}$. Equation (2.10) follows and  as in the
standard case $ J \triangle^{1/2} J  = \triangle^{-1/2}$.

We may then introduce the following hermitean operators
\be
q' \equiv J\,q \,J^* =  i S q S \ ,  \ \ \ \ \ p' \equiv J\,p\,J^*
= - i S p S. \ee They satisfy the following (pseudo-)canonical
commutation relations $$ [q',p'] = J\, [q,p]\, J^* = -i . $$
Furthermore, by the definition of the antilinear operator $S$, we
have $$ (q + i q') \psz = (q - S q S) \psz = 0 \ , \ \ \ \ (p - i
p') \psz = (p - S p S) \psz = 0. $$ The above equations imply
eqs.(2.16). They also imply that the operator
\be
H \equiv (p^2 + {p'}^2)/2 = (p + i p') (p - i p')/2  = (a^* -
b)\,(a - b^*)/2 \ee annihilates $\psz$ and can be taken as the
Hamiltonian, since it has the right commutation relations with $q,
p$. By the same kind of argument, $\psz$  is annihilated also by $
q^2  + (q')^ 2 $. $\blacksquare$

\vspace{1mm}In conclusion, even if the energy spectral condition
is satisfied by the correlation functions of $\omega$, the
corresponding GNS representation of $\pioH$ exhibits features
similar to those of a KMS state. The modular group can be written
as $$(\triangle^{1/2})^{ i t} = e^{- i \pi\, A \,t}, \,\,\,A =
-\ume \,(p q + q p - p' q' - q' p').$$

\vspace{2mm}The representation of  $\Ha$ in an indefinite inner
product space characterized in Proposition 2.3 can be given a
Krein structure.

\begin{Proposition} The indefinite inner product
representation space $D = \pioH) \psz$ can be given a Krein
structure by introducing a metric operator $\eta$ with $\eta \psz
= \psz$ and
\be
\eta a \eta = a \ , \ \ \ \ \ \eta b \eta = - b \ , \ee
equivalently,

\be
\eta q \eta = p' \ , \ \ \ \ \ \eta p \eta = q' \ .
\ee
\end{Proposition}
\Pf \,\,\,\, It follows easily from the positivity of the two
point function $ (\,b\, \psz , \eta\, b\, \psz)$. $\blacksquare$

The so obtained structure provides an example of the
Tomita-Takesaki theory in indefinite inner product spaces;
surprisingly, the commutant is described by pseudo-canonical
variables, as it happens for the time component of the
electromagnetic potential in the Gupta-Bleuler formulation of the
free electromagnetic field. The above representation of $ b, b^*$
is the same as the anti-Fock representation of the CCR discussed
in ~\cite{MSV,MS}

\vspace{2mm} The above representation of the Heisenberg algebra in
a Krein space allows for the construction of the Weyl algebra
(equivalently of the Heisenberg group) as the algebra generated by
the (pseudo) unitary operators  $ U(\alpha) \equiv \exp i \alpha
q$, $  V(\beta) \equiv \exp i \beta p $. Therefore, in this way
one gets a ground state (regular) representation of the Weyl
algebra for the free particle; however, the correlation functions
of ther Weyl operators do not satisfy the energy spectral
condition (see Appendix A).

%%%%%%%%%%%%%%%%%%%%33333333333333333333333333333%%%%%%%%%%%%%%%%%

The lack of positivity of the time translationally invariant state
can be avoided by looking for representations of the Weyl algebra
which do not yield representations of the Heisenberg algebra. The
analog of eq.(2.2) is now \be{\at(W(\a,\b))= W(\a,\b +\a t).}\ee
and the (unique) representation defined by a translationally
invariant state $\Omega$ is identified by the following
expectations
\be
 \O ( W(\alpha, \beta) ) = 0, \ \ \ \ if \,\,\,\, \alpha \neq 0  \ ; \ \ \ \
 \O ( W(0 , \beta) ) = 1 \ .
\ee In this case  the Fourier transforms of the correlation
functions $ \O (A \alpha_t(B))$ are measures with support
contained in $\reali^+$, see Appendix B. The corresponding
euclidean formulation satisfies Nelson positivity and therefore it
admits a unique functional integral representation (see Appendix
C).
%%%%%%%%%%%%%%%%%%%%%%%%%%%%%%%%%%%%%%%%%%%%%%%%%%%%%%%%%%%%%%%%%%
%%%%%%%%%%%%%%%3333333333333333333333333%%%%%%%%%%%%%%%%%%
%%%%%%%%%%%%%%%%%%%%%%%%%%%%%%%%%%%%%%%%%%%%%%%%%%%%%%%%%%%%%%%%%%%%%%%%%%
%\newpage
\section{Euclidean formulation and stochastic processes. Indefinite
case}\vspace{1mm} {\bf a. Schwinger functions and
Osterwalder-Schrader reconstruction}

The n-point Schwinger functions obtained by Laplace transforms of
the Wightman functions of $q$ factorize as usual for Gaussian
states and define a functional $\Phi_S$ on the euclidean Borchers'
algebra, fully determined by the two point function
$$\Phi_S(\overline{f} \times g ) \eqq < f, \,g > \eqq \int
d\t_1\,d\t_2\, \overline{f}(\t_1)\,\S(\t_1, \t_2)\,g(\t_2),\,\,\,
f, g \in \S(\Rbf) \eqq \S,$$ \be{\S(\t_1, \t_2) = S(\t) = c -
|\t|/2.}\ee

Euclidean fields $ x(f), \,f \in \S, \,f $ real, can be introduced
as usual, with two point function  $$< x(f)\, x(g) > = < f, \,g
>$$ and n-point functions given by the  functional ($f$
real) \be{ < e^{i x (f)} > \equiv  e^{ - <f, f>/2} = e^{- \int
f(\t) f(\s) \,S(\t - \s)   d\t \; d s/2}. }\ee The
Oster\-wal\-der-Schra\-der positivity is not satisfied, since
$\S(-\t_1, \t_2) = c - (\t_1 + \t_2)/2, \,\,\t_1, \t_2 \geq 0,$ is
not a positive kernel and therefore $$\Phi_S(\overline{\theta f}
\times f), \,\,\,\, (\theta f)(\t) \eqq f(- \t),$$ is not
positive.

However, the Osterwalder-Schrader (O-S) reconstruction of the real
time indefinite vector space, in terms of the Schwinger functions
can be done by the standard extension of the recostruction without
positivity ~\cite{JS}. The O-S scalar product is defined by
factorization starting from  $$ < f, \, g
>_{OS}  \equiv  < \theta f , g > = i/2
\; (\overline {\tilde f' (0)} \, \tilde g (0) - \overline {\tilde
f (0)} \, \tilde g' (0)) + c \overline {\tilde f (0)} \, \tilde g
(0)  $$ for $f,g \in \S (\reali^+) \eqq \S^+ $.

The null space of the O-S scalar product, $$ N_{OS} \equiv \{  f
\in \S (\reali^+): \, <f,g>_{OS} \, = 0 \ \ \forall g \in \S
(\reali^+) \}  \ \  , $$ has codimension two, so that the space
$\S (\reali^+) / N_{OS}$ is two--dimensional. \vspace{1mm}
%%%%%%%%%%%%%%%%%%%%%%%%%%%%%%%%%%%%%%%%%%%%%%%%%%%%%%%%%%%%
\newline {\bf b. Nelson space} \newline In contrast with the
standard case,  $< . , . >$ is not positive and therefore does not
define a Gaussian measure for the functional integral
representation of the correlation functions (3.2).
\begin{Proposition} The inner product $< . , . >$ is positive
on $\S_0(\Rbf) \eqq \{f \in \S(\Rbf), \,\int f d\t= 0 \}$, and
there is a function $\chi$ such that $< \chi, \, \chi > = -1$.
Thus $\S(\Rbf)$ is an indefinite inner product space with one
negative dimension (pre-Pon\-tri\-a\-gin space).
\end{Proposition}
\Pf \,\,\,The proof follows easily since the Fourier transform of
$S$ is $$\tilde{S}(\o) = 2 \pi c\,\d(\o) - (d/d\o) P(1/\o),$$
where $P$ denotes the principal value. $\blacksquare$

\vspace{2mm} For simplicity, in the following we consider the case
$c=0$. The above indefinite inner product space has a functional
structure very similar to the indefinite inner product space
defined by the (real time) two point function of the massless
scalar field in two space time dimension ~\cite{MPS}. As in that
case, there is only one extension ~\cite{MS} of $\S(\Rbf)$ to a
weakly complete inner product space $\overline{\S}(\Rbf)$,(also
briefly denoted by $\overline{\S}$), since the negative space has
finite dimensions ~\cite{MPS,MS}.

\begin{Proposition} The space $\overline{\S}(\Rbf)$ has the
following decomposition, where $<\oplus>$ denotes orthogonal sum
with respect to $< ., . >$, \be{\overline{\S}(\Rbf) =
\overline{\S}_{00} <\oplus> \V , }\ee $$\S_{00} \eqq (\frac{d}{d
\t})^2 \S^+ <\oplus> (\frac{d}{d \t})^2 \S^-, \,\,\, \S^\pm \eqq
\S(\Rbf^\pm), \,\,  \V \eqq \{a \d_0 + b w,\, \,a, b \in \Cbf \},
$$ \be{ < \d_0, \, \d_0 > = 0 = < w, \, w >, \,\,\,< \d_0, \, w
> = - 1/2, }\ee \be{ < \d_0, \, f > = - \int d \s \,f(\s) |\s|,
\,\,\, < w, \, f >  = - \tilde{f}(0)/2,}\ee the product $< . , .
>$ is positive on $\S_{00}$ \be{< f, \,g > = (F', \,G')_{L^2},
\,\,f= F'', \,g = G'', \, F, \,G \in \S^\pm},\ee and
$\overline{\S}_{00}$ denotes the closure of $\S_{00}$ with respect
to it.

$\overline{\S}(\Rbf)$ can be turned into a Krein space $K_\a$,
such that the above decomposition is also orthogonal with respect
to the Krein scalar product ($\alpha \in {\bf R}^+$) \be{ [ f, \
g ]_\a = < f, \, \eta_\a\, g
> \eqq < f, \,g > + 2 < f, \,\a \d_0 +  \a^{-1} w > < \a \d_0 +
\a^{-1} w, \,g >, }\ee where the metric $\eta_\a,\, \eta_\a^2 = 1$
is defined  by $\a \d_0 + \a^{-1} w$ being its negative
eigenvector.
\end{Proposition}
\Pf \,The sequences $d_n$, with $d_n(\t)$ smooth positive
approximations of the Dirac delta function $\d(\t)$, converge
weakly in the topology of the inner product (3.1); also the
sequences $f_n(\t) = f(\t - n)/n, \, f \in \S(\Rbf),
\,\,\,\tilde{f}(0) = 1$, are weakly convergent. In fact one has
\be{ \lim_{n \ra \infty} < \d_n, \,g > = -
{\scriptstyle{\frac{1}{2}}} \int g(s) |s - \s| d s,\,\,\,\,\,
\lim_{n \ra \pm \infty} < f_n, \,g > = \mp \tilde{g}(0)/ 2.}\ee
Actually, all the above sequences are uniformly weakly compatible
sequences in the sense of ~\cite{MS}, i.e. $\lim_{n, \,m} <u_n,
\,u_m >$ exists and $\lim_n \,< u_n, \,g >$ exists $\forall g \in
\S(\Rbf)$. Therefore they define a weak extension $\S_{ext}$ of
$\S(\Rbf)$ through the addition of the elements \be{ \d_0 =w-
\lim_{n \ra \infty}\,d_n, \,\,\,\,\, w = w- \lim_{n \ra \infty}
\,f_n. }\ee Since $$\lim_{(n,m) \ra \infty}< d_n, d_m
> =0, \,\,\lim_{(n,m) \ra \infty}< f_n,  \d_m
> = - 1/2, \,\, \lim_{(n,m) \ra \infty}<d_n, d_m> = 0$$ we
have \be{< w, \,g > = - \tilde{g}(0)/2,
 \,\,\,< w, \,w > = 0, \,\,\,< \d_0, \, w > = - 1/2, \,\,\, ,<\d_0, \d_0> = 0.}\ee
The product (3.7) is well defined in $\S_{ext}$ and makes it  a
pre-Krein space;  $\S(\Rbf)$ is dense in the Krein completion of
$\S_{ext}$, which therefore coincides with the unique weakly
complete extension  $\overline{S}(\Rbf)$ of $\S(\Rbf)$.

$\overline{S}(\Rbf)$  contains also the weak limits $\d_\s$ of the
sequences  $d_n(\t - \s)$, $\forall \s \in \Rbf$, which satisfy
\be{\,\,\,\,\,< \d_\s, \d_\rho > = - {\scriptstyle{\frac{1}{2}}}
\,|\s - \rho|.}\ee

In order to prove eq.(3.3), we have to show that  $f \in
\overline{\S}, \,\,< f, \, \V > = 0$, and $< f, \, \S_{00}
> =0 $ implies $f = 0$. In fact, if $f_n \ra \,f$ strongly
in $\overline{S}(\Rbf)$, then $<f_n,\d_0> \ra 0$, $<f_n, w> \ra
0$. One can then construct a sequence $h_n \ra f$ with $<h_n, w>
=0 = <h_n, \d>$; by eq.(3.10) this implies $h_n \in \S_0(\Rbf)$,
i.e. $h_n = d\,g_n/d \t$. Then, one has $$ [ h_n - h_m, \, h_n -
h_m ]_\a = < h_n - h_m, \, h_n - h_m
> =  || g_n - g_m ||^2_{L_2}, $$ so that $g_n$ converge in $L_2$. Now,
$< f, \,  \S_{00} > =0$ implies that $\forall k^\pm \in \S^\pm$
$$ < h_n, \, d^2 k^\pm/d\t^2 > = (g_n, \,d k^\pm/d \t)_{L_2} \ra
0,$$ so that $g_n \ra 0,$ in $L_2$ and $h_n \ra 0$ in
$\overline{\S}$.

The other statements of the Proposition follow easily.
$\blacksquare$

\vspace{3mm} In the parametrization of the Krein scalar products,
$\alpha$ plays the r\^{o}le of a scale parameter; thus, even if
the correlation functions are scale invariant, the Krein products
are not.

As usual, one can construct the Fock space $\Gamma(K)$ over $K$.
Euclidean fields  $x(f)$, $f$ real, act on $\Gamma(K)$ and by
standard arguments they can be extended from $\S$ to $K$; the
existence of $\d_0$ and $w$ are equivalent to the existence of the
fields $x(\t)$ at fixed times, $x \eqq x(0)$, and  of the
(positive time) ergodic mean of the velocity, namely $v \eqq
\lim_{\t \ra +\infty} x(\t)/ \t$.

The OS scalar product  extends by continuity to the Krein closure
of $\S^+$ and $K_{OS}$ can be identified with the subspace $\V$ of
the Krein closure $ \overline \S (\reali^+) $ generated by $\d_0$
and $w$. In fact, by eqs. (3.4), (3.5), $\forall f, \,g \in \S^+$
$$ < f, \,g >_{OS} \, = \,< E_0 f, \, E_0 g >, \,\,\,
 E_0 f \eqq \tilde f (0) \, \d_0 + i \tilde f' (0) \, w .$$
%By comparison with eqs. (2.3), one has the correspondence $ x \sim
%q \Psi_0, \,\,\, v \sim i p \Psi_0 $.  Clearly, the one particle
%space determines  the whole (indefinite) OS reconstruction space,
%which coincides with the real time space of Prop. 2.1, with the
%correspondence $$ P(x) \, Q(-i v) \sim P(q) \, Q(p) \, \Psi_0. $$
The above decomposition allows to control the failure of Nelson
positivity and from this point of view the model can be seen as an
example of a general structure which substantially generalizes
Nelson strategy to cases in which positivity fails and in
particular allows for a generalization of the formulation of the
Markov property in terms of projections (see Appendix D).

%%%%%%%%%%%%%%%%%%%%%%%%%%%%%%%%%%%%%%%%%%%%%%%%%%
%%%%%%%%%%%%%44444444444444444444444444444444%%%%%%%%%%%%%%%%%

\section{Functional integral representation}
The decomposition (3.3) provides the tool for writing a functional
integral representation of the functional (3.2). In fact, by
eq.(3.6), $< e^{i x (f)} >, \,\,f \in \S_{00}, \, f= F''$, can be
represented in terms of a functional integral with Wiener measure
\be{< e^{i x (f)} > = \int d\,W_0(\xi(\t)) \, e^{i \xi (f)} , }\ee
since for $f \in \S_{00}$ one has $\int d \,\t |\t| f(\t) = 0 $
and therefore $$ <  f, \, f > = \int d W_0(\xi(\t))\, \xi(f)^2.$$
We are then left with the case $f \in \V$.

Quite generally, a Gaussian functional defined by a (not
necessarily positive) quadratic form on a finite dimensional space
$V$ has a canonical representation in terms of  a positive measure
and  complex random variables.

This is easily seen in our case, where the two dimensional space
$\V$ is generated by $\d_0$ and $w$. Then, one has \be{< \a \d_0 +
\b w, \,\a \d_0 + \b w > = - \a \,\b = 2\pi^{-1}\int d z \,d
\bar{z}\,e^{-2 |z|^2} (\a z - \b \bar{z}) (\a z - \b \bar{z})}\ee
Thus, for any polynomial $\P$ one has \be{ < \P(x, \,v) > =
\pi^{-1} \int \, d z \,d \bar{z}\,e^{-2 |z|^2}\, \P(z,
\,\bar{z}).}\ee In conclusion, the decomposition according to
eq.(3.3) $$\d_\t(\s) = \d_0 + |\t| w + h_\t(\s), \,\,\,h_\t(\s)
\in \overline{\S}_{00}$$ gives the following decomposition of the
field variables \be{ x(\t) = x(\d_\t) = x + |\t| v + x(h_\t),}\ee
with $$< x(h_\t) \,x(h_\s) > = +1/2 (- |\t - \s| + |\t| + |\s|).$$
Therefore, putting $z = z_1 + i z_2$, $$< x(\t_1) ... x(\t_n)>\,
=$$ \be{ 2 \pi^{-1}\int d W_{0,0}(\xi(\t))\, d z_1 \,d z_2 \,e^{-2
|z|^2}\, (\xi(\t_1) + z - |\t_1| \bar{z})...(\xi(\t_n) + z -
|\t_n| \bar{z}).}\ee

The occurrence of complex random variables should not be regarded
as an oddity, since the operators $x, \,v$ are hermitean with
respect  to the indefinite product but they are not self adjoint
operators in the Hilbert-Krein $K_\a$, since they do not commute
with the metric  $\eta_\a$. Their realization as multiplication
operators should therefore account for their non real spectrum and
lead to complex variables. This phenomenon looks of general nature
and therefore should generically appear in local formulations of
gauge theories, where positivity does not hold (for an example see
Ref. ~\cite{LMS3}.

\vspace{2mm} The above functional integral representation gives
direct information on the Hilbert-Krein structure which can be
associated to the correlation functions (4.5). In fact, the Krein
scalar product (3.7) can be associated to  the Krein two point
function of a Gaussian euclidean field $x_\a(f)$ $$[\, f, \, g
\,]_\a = [ x(f) \Psio, \,x(g) \Psio ]_\a = < \Psio,
x_\a(f)\,x_\a(g)\,\Psio > , $$ $$x_\a(f) \eqq 1/2(x(f) + \eta_\a\,
x(f)\, \eta_\a) -i/2 (x(f)  - \eta_\a\, x(f) \,\eta_\a ).$$ The
positive Gaussian functional \be{ [ e^{i x(f)} ]_\alpha \equiv
e^{-  [ f , f ]_\alpha/2},}\ee defines a Gaussian measure on $\S'$
\be{ \int d \mu_\a \,e^{i x(f)} \eqq e^{-  [ f , f ]_\alpha/2}
}\ee and it is not difficult to see that the explicit form of the
corresponding functional integral is obtained  by replacing the
complex variables $z, \,\bar{z}$ in eq.(4.5) by real variables
$\a^{-1} (z_1 + z_2), \,\,-\a( z_1 - z_2)$, corresponding to  $x,
\,v$.

An explicit analysis of the gaussian measure defined by the Krein
scalar product (generating kernel, Markov properties etc.)
immediately follows from such a representation.

The Markov property for $d \mu_\a$ holds in the variables
$\tilde{\xi}(\t) \eqq x + \xi(\t)$, $v(\t)$, with $v(\t)$ constant
in $\t$; in fact, $d \mu_\a$ defines a measure on trajectories
$\tilde{\xi}(\t),\,v(\t)$, concentrated on $v(\t) = v$ and
satisfying the Markov property in the two variables $x(\t),
\,v(\t)$.

Clearly, the process defined by $d \mu_\a$ corresponds to Brownian
motion, with gaussian distribution of variance $\alpha^{-2}/2$ for
the position at $\t=0$, with an additional gaussian variable, with
variance $\alpha^2/2$, describing non-zero mean velocities, with
opposite and constant values for $\t > 0$ and $\t<0$.  The limit
$\alpha \to 0$ exists in the sense of positive functionals on the
Bohr algebra and coincides with the measure on its spectrum
discussed in Appendix C.

%%%%%%%%%%%%%%%%%AAAAAAAAAAAAAAAAAAA%%%%%%%%%%%%%%%%%%%%%%%%%%%%%%%%%%%%%%%
%%%%%%%%%%%%%%%%%%%%%%%%%%%%%%%%%%%%%%%%%%%%%%%%%%%%%%%%%
\newpage
\appendix\section{Appendix A. Indefinite representations of the Weyl algebra}
\begin{Proposition} The formal series corresponding to the
exponentials $ U(\alpha) $, $ V(\beta) $, $\alpha, \beta \in
\reali $ converge strongly in the Krein topology defined in
Proposition 2.4 and define (pseudo) unitary operators in the
corresponding Krein space. The so obtained Weyl operators $
U(\alpha)$, $ V(\beta) $ generate a Weyl algebra $\Wa$ and in this
way one gets a regular representation in a Krein space, defined by
a state $\o$ invariant under the free time evolution and
characterized by the following expectations
\be
\o ( e^{i \alpha q} \, e^{i \beta p} ) = e^{-i \alpha \beta /2 } \
\ \ \ \ \forall \alpha, \beta \in \reali \ee
\end{Proposition}
\Pf \,\,\,\,  By expressing $q$ and $p$ in terms of the
destruction and creation operators  $a, a^*, b, b^* $  defined in
Proposition 2.3, one gets (by Fock space methods) the analog of
the standard Fock estimate $$
  || q^n \psz ||_K \leq 2^n \sqrt {n!},
$$ where $ || \ ||_K$  denotes the Krein-Hilbert norm, and a
similar estimate for $ p^n \psz $. As in the standard case, this
yields the strong convergence of the series and the existence of
the (pseudo) unitary operators $ U(\alpha) $, $  V(\beta)$. The
above expectations follow from the correlation functions
\be
\o(q^n p^m) = \delta_{n,m} (i/2)^n \, n! \ee

It is worthwhile to mention that the operators $ U(\alpha)$, $
V(\beta)$ do not commute  with the metric operator $\eta$, and
therefore they are not unitary with respect to the positive
(Hilbert) scalar product $ ( \cdot , \eta \cdot ) $ defined by
$\eta$. Since $\o$ is time translation invariant, the time
evolution automorphisms $\alpha_t$, $t \in \reali$  are
implemented by a one parameter group $\Ut $ of (pseudo) unitary
operators, and information about their spectral properties is
given by the Fourier transforms of the correlation functions $\o
(A \,\alpha_t(B))$.

\begin{Proposition} In the representation of the Weyl algebra in
the Krein space discussed above, the Fourier transform of the
correlation functions $ \o (A \,\alpha_t(B))$ are tempered
distributions with the following properties:

\ni i) for $ A, B \in \Ha $ they have support at the origin,

\ni ii) for $ A, B \in \Wa $ they are measures with support in the
real line.
\end{Proposition}
\Pf \,\,\, In fact, by the definition of the time evolution and
the invariance of $\o$ one has $$
 \o (p^k q^j \, \alpha_t( p^l q^m)) = P_m (t),
$$ with $ P_m(t) $  a polynomial of degree  $m$. On the other hand
$$
 \o (U(\alpha)  \,\alpha_t ( U(\beta ))) = e^{-i \alpha \beta t/2}.
$$ In the latter case, the support of the Fourier transform is
contained in the positive real axis for the diagonal expectations,
$\alpha = - \beta$, but not in general.

%%%%%%%%%%%%%%%%%%%%%%%%%%%%%%%%%%%%%%%%%%%%%%%%%%%%%%
%%%%%%%%%%%%%%%%%%%%BBBBBBBBBBBBBBBBBBBBB%%%%%%%%%%%%%%%%%%%%%%%%%%%%%%%%%%%
\section{Appendix B. Ground state positive representations of
the Weyl algebra}

\begin{Proposition} A time translationally invariant positive
state $\O$  on the Weyl algebra $\Wa$ satisfying the positivity of
the energy spectrum is identified by  having the following
expectations of $ W(\alpha, \beta) \equiv U(\alpha) V(\beta) \exp
(i \alpha \beta /2)$
\be
 \O ( W(\alpha, \beta) ) = 0, \ \ \ \ if \,\,\,\, \alpha \neq 0  \ ; \ \ \ \
 \O ( W(0 , \beta) ) = 1 \ .
\ee
\end{Proposition}
\def \at {\alpha_t}
\Pf\,\,\,\,\,  Time translation invariance implies that the above
expectation is independent of $\beta$ if $\alpha \neq 0$. On the
other hand, \be{\Omega(W(\a,0)\,\at(W(\g,0)) = \Omega(W(\a + \g,\g
t))\,e^{-i \a \g t/2},}\ee so that, since for $\a = \g \neq 0$,
$$\Omega(W(\a + \g, \g t)) = \Omega(W(\a+\g, 0)),$$ positivity of
the energy requires that it vanishes. For $\a = - \g$, the Fourier
transform of eq.(B.2) has support in $\Rbf^+, \, \forall \g \in
\Rbf$,  iff the Fourier transform of $\Omega(W(0, \g t))$ has
positive support, which holds  for all $\g$ iff the support is at
the origin. Positivity of the state $\Omega$ then implies that
$\Omega(W(0, \a t)) = 1$. Thus, eqs.(B.1) hold. Conversely,
eqs.(B.1) define a positive state (as a limit of ground states of
harmonic oscillators ~\cite{AMS1}). Moreover, eqs.(2.2) imply
$$\Omega(W(\a, \b)\, \at(W(\g, \d))) = \d_{-\a, \g}\,e^{- i \a(\d
+ \b)/2}\,e^{i \a^2 t/2},$$ where $\d_{\a, \gamma} = 1$, if $\a =
\gamma$ and zero otherwise; therefore positivity of the energy
follows.

\vspace{2mm} In conclusion, by the GNS construction, one has a
nonregular representation of the Weyl algebra in a Hilbert space
$\H$, with cyclic vector $\Psi_{\Omega}$. Since eqs.(B.1) imply
\be{V(\b) \,\Psi_{\Omega} = \Psi_{\Omega},}\ee $\Psi_\Omega$ is
also cyclic with respect to the algebra generated by the
$U(\a)$'s. The occurrence of non regular representations should
not be regarded as too bizarre, since they can be related to
reasonable physical descriptions. In fact, if we consider a free
particle in a bounded volume $V$, it is reasonable to consider the
algebra of canonical variables $\AV$ generated by $\exp (i \beta
p)$, $\b \in \reali$ and by the (continuous) functions of $q$,
$f_V(q)$,  with support contained in $V$.

There is a natural embedding of $\AV$ into the Weyl algebra $\Wa$,
i.e., if $$ f_V (x) = \sum c_n \, e^{i k_n x} \ , \ \ \ \ \ x \in
V , $$ then its periodic extension $$ f (q) =  \sum c_n \, e^{i
k_n q} \ , \ \ \ \ \ q \in \reali, $$ defines a corresponding
element of $\Wa$. Then, if $\psV (x) = const$ denotes the ground
state in the volume $V$ (with periodic or Neumann boundary
conditions for the Hamiltonian), one has $$ (\psV , f_V (x)  \,
e^{i\,\b\,p}\, \psV) = \O(f(q) \,e^{i\,\b\,p}), $$ where $\O$ is
the nonregular state characterized in the above Proposition. The
nonregular representation provides therefore a volume independent
mathematical description of the above concrete situation.

%%%%%%%%%%%%CCCCCCCCCCCCCCCCCCCCCCCCCCCCCCCCCCCCCCCCCCCCC%%%%
%%%%%%%%%%%%%%%%%%%%%%%%%%%%%%%%%%%%%%%%%%%%%%%%%%%%%%%%%%%%%%%%%

\section{Appendix C. Positive euclidean formulation and stochastic processes}
In order to discuss the stochastic processes associated to
the quantum free particle, it is convenient to derive the
corresponding euclidean formulation.
\def \a {\alpha}

For the representation characterized in the previous Appendix, we
have for the two point (Wightman) function for the Weyl operators
$U(\a)$ \be{ \O( U(-\a)\,e^{i H t}\,U(\a')) = \d_{\a, \a'}\, e^{i
\a^2\,t/2}.}\ee  The Fourier transform is $$\tilde{W}(\o) =
\sqrt{2 \pi} \,\d(\o - \a^2/2)\,\d_{\a,  \a'}.$$ The n-point
(Wightman)functions of the Weyl operators $U(\a)$ are obtained by
induction from $$\at(U(\a)\,V(\b))= U(\a)\,V(\b + \a t)\,e^{i \a^2
t/2}$$ using eq.(C.1), and are given by ($U(\gamma, t) \eqq
\alpha_t (U(\gamma)))$

\be{\Omega((U(\alpha_1, t_1))...U(\alpha_n,t_n))) =
 \delta_{\sum
\alpha_i,\,0}\,e^{i \sum_{i=2}^n (t_i - t_{i-1}) (\sum_{k=i}^n
\alpha_k)^2/2}.}\ee The corresponding n-point Schwinger functions
are immediately obtained by analytic continuation to ordered
imaginary times $\t_1\leq\t_2 ...\leq\t_n$ : \be{\S(\a_1 \t_1,
...\a_n \t_n)=\delta_{\sum \alpha_i,\,0}\,e^{- \sum_{i=2}^n (\t_i-
\t_{i-1}) (\sum_{k=i}^n\alpha_k)^2/2}.}\ee and extended by
symmetry to all euclidean times $\t_1, \t_2 ...\t_n \in \Rbf$.

>From the existence and positivity of the Hamiltonian in the
representation defined by eq.(B.1), it follows that the above
Schwinger functions can also be written as \be{\S(\a_1 \t_1,
...\a_n \t_n)= (\Psi_\Omega, U(\a_1) e^{-(\t_2-\t_1)H}U(\a_2)
...e^{-(\t_n-\t_{n-1})H} U(\a_n) \Psi_\Omega)}\ee By standard
arguments, one can introduce the corresponding Borchers algebra
and euclidean fields $U_E(\a,\t)$ so that the Schwinger functions
define a linear functional $E$ on the euclidean fields $$\S(\a_1
\t_1, ...\a_n \t_n)= E(U_E(\a_1,\t_1)...U_E(\a_n, \t_n)).$$

Equation (4.3) implies the Osterwalder-Schrader (OS) positivity of
the Schwinger functions, with the OS reflection operator $\theta$
defined by $$\theta U_E(\a, \t) = U_E(\a, -\t),$$ i.e. one has
$$E(\overline{\theta B}\,B) \geq 0,$$ $\forall B$ belonging to the
algebra generated by $U_E(\a, \t),\, \a\in \Rbf, \,\t \geq 0$,
where $\overline{U_E(\a,\t)} \eqq U_E(-\a,\t)$.

It is a non trivial fact that also Nelson positivity holds. As a
matter of fact, the above Schwinger functions can be expressed as
expectations of fields $ e^{i \a x(\t)}$ with functional measure
$$d\mu(x(\t)) = dW_{0,x}(x(\t)) \,d\nu(x),$$ where $dW_{s,y}$
denotes the Wiener measure for trajectories starting at the point
$y$ at time $\t=s$ and $d\nu$ denotes the ergodic mean; in fact $d
\nu$ defines a measure on the Gelfand spectrum $ \Sigma$ of the
Bohr algebra generated by  $ e^{i \a x}$ ~\cite{LMS1} and $$\int d
\nu(x) \,d W_{0,x}(x(\t)) \,e^{i\a_1 x(\t_1)}...e^{i\a_n
x(\t_n)}$$ $$= \int d \nu(x) \, \int d W_{0,0}(y(\t)) \,e^{i
\sum_i \a_i \,y(\t_i)}\,e^{i \sum_i \a_i \, x}$$ $$= \d( \sum_i
\a_i )\,\int d W_{0,0}(y(\t)) \,e^{i \sum_i \a_i y(\t_i)} =
\S(\a_1 \t_1, ...\a_n \t_n), \,\,\,\,y(\t_i) \eqq x(\t_i) - x.$$

The  measure $d \mu$ is invariant under time translations, $$d
W_{0,y}(x(\t)) d \nu(y) = d W_{s,y}(x(\t)) d \nu(y)$$ because $d
\nu(x)$ is invariant under translations $x \ra x + a$ and
therefore stationary for the Brownian motion.

The measure $d \mu$ is invariant under translations $x(\t) \ra
x(\t) + a $; they can be given the meaning of gauge trasformations
and can be regarded as the analog of the gauge group of
traslations of step $2 \pi$ for a particle on a circle
~\cite{LMS1}. In conclusion, the euclidean correlation functions
can be obtained as the stochastic process with expectations
defined by $d\mu$.

Since the ground state $\Psi_\O$ is cyclic with respect to the
euclidean algebra at time zero in the space $\H$, obtained by the
OS recostruction theorem, such a space can be identified with
$L^2(\Sigma, d\nu)$.

One can explicitly check that the Markov property holds; if $\t_1
\leq 0 \leq \t_2$ one has $$\int d \mu\, e^{i\a x(\t_1)} \,e^{i\b
x(\t_2)}=\int d \nu(x) e^{i (\a+\b) x} \int d W_{0,x}(x(\t))
e^{i\a (x(\t_1)-x)} e^{i\b (x(\t_2)-x)}$$ $$=\int d \nu(x)
e^{i(\a+\b)x)} d W_{0,0}^-(y(\t))\, e^{i \a y(\t_1)}\, d
W_{0,0}^+(y(\t))\, e^{i \b y(\t_2)}.$$

%%%%%%DDDDDDDDDDDDDDDDDDDDDDDDDDDDDDDDDDDDDD%%%%%%%%%%%%%%%%%%%%%%
%%%%%%%%%%%%%%%%%%%%%%%%%%%%%%%%%%%%%%%%%%%%%%%%%%%%%%%%%%%%%%%%%
\sloppy
\section{Appendix D. Markov property without Nelson positivity}
\fussy
 For simplicity, we discuss the problem at the level of the
two point function, which is assumed to define a non degenerate
inner product $< . \,, .
>$ on $\S$, satisfying  the following properties:
\vspace{1mm}\newline i) $< \theta f, \, \theta g > = < f, \,g
>, \,\,\,\, (\theta f)(\t) \eqq f(-\t),$
\vspace{1mm}\newline ii) there exists an operator $D$ on $\S$,
such that, $\forall f, g \in \S$ $$< f, \,D g > = ( f , g)_{L^2},
\,\,\,D\S^\pm \subseteq \S^\pm,\,\,\,\,\,[\,D, \, \theta\,] = 0,$$
iii) $\S$ is weakly dense in a nondegenerate weakly complete inner
product space $\overline{\S}$, which has the following
decomposition \be{ \overline{\S} = \overline{\S^-} <+>
\overline{D\S^+},}\ee with $<+>$ denoting a $< , >$ orthogonal
sum.

\def \bSp {\overline{\S^+}}
\def \bSm  {\overline{\S^-}}
\def \bSpm {\overline{\S^\pm}}
\def \bDSpm {\overline{D\S^\pm}}
\def \bDSm  {\overline{D\S^-}}
\vspace{1mm}Then, by i), ii) one also has \be{  \overline{\S} =
\overline{\S^+} <+> \overline{D\S^-}.}\ee Furthermore, since $
\bDSpm \subseteq \bSpm$, $\forall f \in \S$, by  eq.(D.1) one has
$$f = f_- + f_{0+}, \,\,\,\,f_- \in \bSm, \,\,\,\,\,f_{0+} \in
\overline{D\S^+}$$ and by eq.(D.2) $f_- = (f_-)_+ + (f_-)_{0-}, \,
\,\,(f_-)_+ \in \bSp, \,\, (f_-)_{0-} \in \bDSm$. Hence, $$ f =
f_{0+} + (f_-)_{0-} + (f_-)_+, \,\,\,\, (f_-)_+ \in \bSp \cap
\bSm.$$
%Moreover, if $f \in \bSp\cap\bSm$, then $f_{0+} =0 =
%(f_-)_{0-}$, i.e. $f=(f_-)_+$.
In conclusion, one has \be{
\overline{\S} = \overline{D\S^-} <+> \overline{D\S^+} <+> \V,
\,\,\,\,\,\,\,\,\,\V = \overline{\S^+} \cap \overline{\S^-}.}\ee
Since  $\bar{\S}$ is non degenerate the decompositions (D.1-D.3)
are unique; in fact, non uniqueness would imply a non zero
intersection of orthogonal spaces, which implies degeneracy of the
inner product. Thus, any vector $f \in \bar{\S}$ has unique
decompositions according to eqs.(D.1-D.3), which means that
correspondingly there are everywhere defined idempotent operators,
which are hermitean with respect to $ < ,
>$. In particular, there are $E_\pm, \,E_0$ defined by $E_\pm
\overline{\S} = \overline{\S^\pm}$, and $E_0 \bar{\S} = \V$ and
eqs.(D.1-D.3)  give \be{E_+ E_- = E_- E_+ = E_0.}\ee Thus, we get
the same operator formulation of the Markov property as in the
positive case.

For quasi free states, the above construction extends in the usual
way to the n-point Schwinger functions, in terms of the sum of the
symmetric tensor products of the above indefinite spaces.
Similarly, the euclidean field $x(f)$ has an extension to $f \in
\bar{\S}$.

In the example discussed in this note, $D = d^2/ d \t ^2$ with the
Krein structure  given by eq.(3.7). The decomposition (D.3)
reduces to (3.3) and it is also  orthogonal in the positive scalar
product. The lesson from the model is that the space $\V$ can be
larger than the standard time zero space; in the model it contains
a time translationally invariant variable, actually the (positive
time) ergodic limit of the velocity.

Thus, the model supports the idea that, in the (indefinite) space
defined by the correlation functions of the field algebra,  the
observable algebra has in general a reducible vacuum
representation ~\cite{MPS1, S, LMS3}.

\vspace{30mm} \noindent Keywords: Weyl algebra, non-regular
representations, Heisenberg algebra, indefinite representations,
euclidean quantum mechanics, complex gaussian processes
\vspace{10mm} \newline MSC: 81S40, \,60G10, \, 60G15, \, 46C20, \,
47B50
\end{document}